\documentclass[twocolumn,showpacs,preprintnumbers,prl]{revtex4}

\usepackage{amssymb}
\usepackage{graphicx}

\newcommand{\V}{V}
\newcommand{\N}{{\cal N}}
\newcommand{\B}{{\cal B}}

\begin{document}

\title{Scenario for Spin Glass Phase with Infinitely Many States}
\author{Olivia L. White and Daniel S. Fisher}
\affiliation{Harvard University, Cambridge, MA 02138}
\date{\today}

\begin{abstract}
A possible phase in short-range spin glasses exhibiting infinitely many equilibrium states is proposed and characterized in real space. Experimental signatures in equilibrating systems measured with scanning probes are discussed. Some models with correlations in their exchange interactions are argued to exhibit this phase. Questions are raised about more realistic models and related issues. 
\end{abstract}

\maketitle

The term ``spin glass'' refers to both experimental systems and theoretical models with enough randomness and frustration to preclude conventional magnetic order \cite{BinderYoungFisherHertz}. Despite thirty years of study, the nature of possible spin-glass phases remains controversial. Even the most studied model, the Edwards-Anderson (EA) Ising model with $\mathcal{H} = - \sum_{<i,j>} J_{ij}\sigma_i \sigma_j$ with $\sigma_i = \pm 1$  and short-range couplings $J_{ij}$ independently drawn from a symmetric distribution \cite{EA}, is still not understood. The droplet/scaling scenario for the spin-glass phase has only one pair of states and no known inconsistencies \cite{AndersonPondBMMcMillanFH1,FH2}. Parisi has hypothesized the existence of infinitely many equilibrium states as found in mean-field spin glasses  \cite{Parisi_a}. But in finite-dimensional systems the nature of and relationship between such putative states remain obscure. 

In this paper, we introduce the {\it montage phase}, a concrete scenario for a spin glass phase with many {\it incongruent} states, states differing by other than symmetry in most of the system \cite{FH_many_states}. By characterizing this phase and its experimental signatures, we also generate questions about more general short-range spin glasses. Since real systems do not have the independence and special statistical symmetry of EA models, we consider a broader class of spin-glass models. For illustration, we construct special models that exhibit infinitely many incongruent states at low temperatures.

For experiments it is essential to address a fundamental question: How can one determine --- even in principle --- the number of equilibrium states in a system too large to equilibrate fully? An ideal experiment would map repeatedly the time evolution after quenches of spin correlations in the {\it same} large region of size $\ell$. After a sufficiently long waiting time, a map of local magnetizations in the region will resemble one of the equilibrium states. 
The number of states distinguishable in the region that can arise from random initial conditions gives the number of distinct such maps, $\N(\ell)$. If there are infinitely many incongruent equilibrium states, $\N(\ell)$ will increase with $\ell$. 

The nature of ground states and their excitations usually determines the equilibrium behavior of low temperature phases. Infinite-system ground states are configurations whose energy cannot be lowered by any {\it finite} change. These can be constructed using sequences of fixed boundary conditions on a set of nested boxes, $\{ \B_L \}$, to generate sequences of finite-region ground states. Each such sequence that converges yields an infinite system ground state. At non-zero temperature, a similar procedure yields infinite system equilibrium states, pure states. While most sequences will not converge, any sequence has at least one convergent subsequence \cite{RuelleSinai}. Motivated by Newman and Stein \cite{NS4_groundstates}, on whose work we build, we call the distribution of states that appear in the ensemble of sequences of random boundary conditions the {\it equilbrium metastate}. With short-range interactions, the number of  possible boundary conditions on a box is exponential in its surface area so the number of equilibrium states distinguishable in a box of size $\ell$ grows no faster than $e^{c\ell^{d-1}}$ \cite{Thouless}.

The physics of equilibration from random initial conditions bears some resemblance to formal limits of boundary condition sequences. After a quench, the length scale, $L_w$, over which the system is in local equilibrium increases with waiting time, $t_w$. After times larger than $t_w$,  regions smaller than $L_w$  resemble physical equilibrium states. More precisely, in scale-$\ell$ subregions with $\ell \ll L_w$, correlations averaged over times short compared to $t_w$ are approximated by those of {\it one} of the infinite system equilibrium states \cite{Middleton1,NS1_groundstates,ellscale}. The particular infinite system state that appears depends upon the surroundings of the scale-$L_w$ region, which itself resembles an equilibrated finite-sized system with effective boundary conditions imposed by these surroundings. The {\it maturation metastate} gives the distribution of equilibrium states that appear over the ensemble of all possible initial conditions and ensuing long-time dynamical histories. 

In a ferromagnet, for example, from random initial conditions the $\uparrow$ and $\downarrow$ states appear equally frequently, but which occurs in a given region varies stochastically with $t_w$ as domain walls pass through. Typical scale-$L_w$, regions will contain at most a few domain walls --- no more than boundary conditions on such regions can induce in equilibrium. In principle, there are also equilibrium states with a fixed flat infinite domain wall, but these arise only from special initial conditions or special sequences of boundary conditions and thus appear with zero probability in the metastates.  In any case, domain-wall states differ from both the $\uparrow$ and $\downarrow$ states only in a negligible fraction of the system and thus are {\it regionally congruent} \cite{FH_many_states} to these basic states. 

The equilibrium and maturation metastates of random Ising systems can contain more than two states. A three-dimensional system with all ferromagnetic couplings except for couplings of random sign across a single plane through the origin provides a simple example. The coupling between half-spaces scales as $\sqrt{L^2}$ in an $L\times L\times L$ box centered at the origin so boundary conditions can control each half-space {\it separately}. The resulting {\it four} ground states are regionally congruent because almost every region is the same in each up to the global symmetry. The metastates contain all four states with equal weight since the effects of random boundary conditions and of random couplings are comparable. The balance between these determines the orientations of each half space in any given finite system or at any given time.

A system with many {\it incongruent} states would, in contrast to one with only domain wall states, exhibit multiple locally-distinguishable spin maps for any large $t_w $\cite{FH2,FH_many_states,NS1_groundstates,NS5_groundstates}.
Can this occur in short-range Ising spin-glass models? 

It is instructive to show first that some finite dimensional Ising models can have infinitely many incongruent states. This can occur if there are many infinite clusters of spins that are separately controllable. 

Consider a $d$-dimensional lattice consisting of a stack of slabs of thickness $s$ and dimension $d_c = d-1$. Couplings within each slab are ferromagnetic, while those between them --- across borders of dimension $\bar{d_c}=d-1$ --- have random signs. The energy to change the orientation of a size $\ell$ region within an isolated slab scales as $\ell^{\theta_c}$ with $\theta_c=d-2$, the cluster stiffness exponent. Inter-slab random coupling scales as $\ell^{\phi_c}$ with $\phi_c=\bar{d_c}/2$. An Imry-Ma argument \cite{ImryMa} implies that each slab is ferromagnetic provided $d-2>(d-1)/2$. Boundary conditions can control the $\mathcal{O}(\ell)$ slabs in box $\B_\ell$ independently. Thus there are $\N(\ell) = \exp(\ell \log 2/s)$  possible slab orientations, each corresponding to an infinite system state. Two randomly chosen ground states are incongruent since most sites are near a slab with different relative orientation in the two states.  Isotropic versions of this model can be made from interpenetrating arrays of randomly coupled $d_c$-dimensional ferromagnetic slabs perpendicular to each set of $d-d_c$ directions. For $d_c>2$, these will have exponentially many incongruent equilibrium states with $\log[\N(\ell)] = \mathcal{O}(\ell^{d-d_c})$. Although they themselves are not spin glasses because both their couplings and spatially averaged spin correlations exhibit long-range order, the slab models suggest an analogous spin-glass phase.

The following fundamental property characterizes a putative spin-glass phase with infinitely many incongruent states which we call a montage phase: Repeated quenches from high temperature reveal clusters of spins with the same relative orientation at long times after each quench. The maturation metastate characterizes the set of such clusters of spins with similar intra-cluster orientation together with the variation of relative cluster orientations with wait time and from quench-to-quench.

To exhibit a montage phase, the system must be composed of infinitely many infinite clusters of spins, each separately controllable by  boundary conditions. Different states will be incongruent only if most spins are close to cluster boundaries, thus (i) together the clusters occupy a finite fraction of space and (ii) their boundary and full dimensions are equal, $\bar{d_c}=d_c$. Clusters will be independently controllable if (iii) in isolation each has a unique pair of ground states, and (iv) they interact with each other weakly enough. The number of distinguishable states will grow with scale as $\N(\ell)\sim \exp(c\ell^{d-d_c})$.

As with the slab models, conditions (iii) and (iv) hold only if the minimal energy cost to invert the spins in a typical scale-$\ell$ section of an isolated cluster is much larger than the typical interaction energy of this section with spins in all other clusters. If the stiffness scales as $\ell^{\theta_c}$ and the interactions as $\ell^{\phi_c}$, this requires 
\begin{equation}
\theta_c>\phi_c\ .  \label{stiffness}
\end{equation}

The stiffness exponent, $\theta_c$, depends on cluster topology and on the range and degree of frustration of the intra-cluster couplings. For any model with nearest-neighbor or sufficiently short-range interactions, $d_c>d-1$ so that clusters span the system and
\begin{equation}
\theta_c \leq d_c -1.  \label{intra_exp}
\end{equation}
More precisely, $\theta_c \leq d_c^|$, the {\it cutting dimension} of a cluster: a minimal cut to sever a typical scale $L$-section from the rest of the cluster breaks of order $L^{d_c^|}$ bonds.  In general, $d_c^|\le d_c-1$, the upper bound obtaining only if the clusters are very well connected. The stiffness exponent will be equal to $d_c^|$ only if bonds within a cluster are sufficiently unfrustrated. 

The interaction exponent $\phi_c$ depends on inter-cluster couplings. For independently-random couplings between clusters when short-range couplings dominate, 
\begin{equation}
\phi_c = d_c/2.   \label{inter_exp}
\end{equation}
Anti-correlations in the inter-cluster couplings that compensate for correlations among intra-cluster couplings could reduce $\phi_c$.

The montage phase is stable to a small uniform (or random) field provided that $\theta_c>d_c/2$.  This follows by analogy to random-field ferromagnets.  In a magnetic field the orientation of each cluster in a finite size system with random boundary conditions will be determined by the competition between the field and the inter-cluster energies.

Can this spin glass phase with infinitely many states exist in EA models?  Can it exist in other short-range Ising spin-glass models that are difficult to distinguish from those with independently random couplings? We first address the latter question by constructing examples.

If interactions within clusters decay as a power of separation, 
\begin{equation}
J_{ij} = \frac{K_{ij}}{|i-j|^{d+\zeta}}
\label{interactions}
\end{equation}
with the $K_{ij}$ of order $1$, cluster stiffness can be substantial even for sparse clusters and is largest for unfrustrated couplings within clusters. The  energy to flip a typical scale-$L$ region of an isolated cluster from its ground state orientation determines $\theta_c$.  Since flipping a typical spin in an isolated unfrustrated cluster costs $L^{d_c}/L^{d+\zeta}$, summing over pairs of spins yields
\begin{equation}
\theta_c = 2d_c - d - \zeta \
\label{theta}
\end{equation}
for $\zeta < 2d_c - d$.

If the strength of inter-cluster interactions also decays with power law Eq. \ref{interactions} but with independently random signs, $K^{\rm inter}_{ij}=\pm \epsilon$ with equal probability, the inter-cluster part of the interaction per spin will be finite as long as $\zeta>-d/2$. Then $\phi_c=d_c/2$ as in the short-range case. Thus such a power-law model exhibits exponentially many incongruent states if $3d_c/2>d+\zeta>d/2$, which is 
satisfiable if $d_c>d/3$. 

When can space-filling sets of fractal clusters exist?  They certainly can be constructed hierarchically. For example, divide a $d$-dimensional (hyper)cube  into $2^{qd}$ smaller cubes and label half of these A and half B. Repeat this process inductively, subdividing each subcube and appending additional labels A and B. Each string of As and Bs will then label a fractal cluster of dimension $d_c = d - 1/q$. Furthermore, provided that $d \geq 3$ and $q$ is sufficiently large, this can be done with some randomness in the labeling but so that nevertheless each cluster is connected and system spanning with boundary dimension $\bar{d_c}=d_c$ and cutting dimension $d_c^|>d_c/2$.

These hierarchically constructed clusters with nearest neighbor interactions that are unfrustrated within clusters and random in sign between them are separately controllable. In this case, $\theta_c=d_c^|> d_c/2=\phi_c$ so that nearest neighbor couplings are sufficient to generate exponentially many states in three (or more) dimensions. A set of signs labels each state-- one sign for every infinite string of As and Bs \cite {footnote}.

More realistic models require space-filling sets of appropriate fractal clusters that are statistically translationally invariant. Do these exist?  Following Newman and Stein \cite{NS1_highly_disordered}, one possibility is built from {\it invasion} percolation clusters \cite{incip-perc}. In dimensions $d>8$, the number of disjoint invasion clusters is believed to grow as a power of the scale.  Similarly, loop-erased random walks in $d>4$ form sets of clusters with $d_c\ge2$ that are collectively space filling \cite{Pemantle1Benjamini1}. 
With power law interactions with special correlations, such clusters can be separately controllable. With short-range interactions, however, it is not known whether they will be, even in high dimensions.

The special interactions in the above power-law example are physically unrealistic. It seems unlikely that conventional EA models exhibit separately controllable clusters because of the apparent need for many weakly frustrated sets. In real spin glasses, however, couplings are neither symmetrically distributed nor independent of other nearby couplings, and can be long range: e.g. RKKY  ($\sim1/r^3$). Can clusters satisfying conditions (i)-(iv) arise spontaneously in systems with interactions that have translationally invariant correlations among them? If so, how short-ranged can the interactions and correlations be? In particular, can the correlations among the couplings come from an effective Boltzmann-like probability measure with only short range interactions in the effective Hamiltonian?

XY models with infinite uniaxial anisotropy suggest some reason for optimism: they both illustrate a possible mechanism for generating many states as well as provide an example in which controllable fractal clusters arise spontaneously. In dimensions greater than four, XY models with random uniaxial anisotropy are believed to exhibit ferromagnetic order \cite{Pelcovits}. When anisotropy is infinite, the random axis at site $i$, ${\bf \hat{a}}_i$, dictates the direction of the XY spin: ${\bf  S}_i = \sigma_i {\bf \hat{a}}_i$ with $\sigma_i = \pm 1$. The resulting Hamiltonian is equivalent to an Ising Hamiltonian with $J_{ij}\propto {\bf \hat{a}}_i\cdot {\bf \hat{a}}_j$. In high dimensions, this model should exhibit ferromagnetic order in the XY variables provided that the coordination number is sufficiently high or correlation in random axis directions is sufficient \cite{FisherXY}. Consequently the direction of the spontaneous magnetization, ${\bf \mu}$, should parametrize an infinite family of states --- with $\sigma_i {\bf \hat{a}}_i \cdot {\bf \mu} > 0$ at most sites --- in a way which would be hidden in the Ising variables.

Although this may be a qualitatively distinct scenario for many states, 
the number of which probably would grow only as a power of length scale {\it some} separately controllable fractal clusters do exist in such models. All spins in a given finite box flip when fixed boundary conditions inducing a spontaneous magnetization $\mu$ are rotated by $\pi$. There must be some much smaller change in boundary conditions inducing a new state, $\mu'$ that is as close as possible to $\mu$ while still differing in the deep interior of the box. Reversal of a connected fractal set of spins that boundary conditions can control relates states $\mu$ and $\mu'$. But interactions between such sets are too strong for this system to be in the montage phase. We expect that the number of controllable fractal sets only grows as a power of system size and that at most only a few of them can be controlled simultaneously.

We now turn to experimentally observable consequences of the montage phase scenario. Maps of local magnetizations (or other local properties) in the same region after repeated quenches --- to the same conditions --- followed by waits for time $t_w$ give the spectrum of states in the maturation metastate as well as their local structure.  Soon after a quench from high temperature, only nearby spins are correlated and there will be $e^{\ell^d/\V}$ distinct possible sets of spin orientations. With increasing $t_w$, $\V$ will increase, and the number of observed maps will decrease. If there are only two states, after a sufficiently long time such maps will coincide in the observed region up to a global spin inversion. But in the montage phase, even after extremely long times, the number of distinct states observed in a scale-$\ell$ region will grow with $\ell$ as 
\begin{equation}
\N(\ell)\sim\exp(\ell^\sigma)
\end{equation}
presumably with $\sigma=d-d_c$. Though estimating the number of realizable states may be impracticable, experiments should be able to determine whether $\sigma > 0$.

Correlations between runs will carry information about the {\it structure} of physical states. Long-distance correlations between pairs of spins on the same cluster, $\langle S_i S_j \rangle$, averaged over a long time (but short compared to $t_w$), will be reproducible from run-to-run whereas the sign of correlations between spins on different clusters will vary. 
So intra-cluster correlations will dominate the mean-square correlations obtained by first averaging over many runs to obtain $\Gamma_{xy}\equiv \overline{\langle S_xS_y\rangle}$ and then averaging $\Gamma_{xy}^2$ over many pairs of spins distance $\ell$ apart. These correlations decay as $\ell^{-(d-d_c)}$ up to the length scale $L_w(t_w)$ on which equilibrium is established. Decay is exponential for $\ell \gg L_w$. Spins in the same cluster have $\Gamma_{xy}=O(1)$.

There are subtleties in determining whether local equilibrium has been reached on a sufficiently large scale $L_w$ relative to $\ell$.  First, pairs of spins on neighboring clusters may appear to be on the same cluster if these clusters tend to have the same mutual orientations at early times. Second, two sites on the same cluster may appear as if they are on distinct clusters if they are linked primarily via other spins that that extend well beyond $L_w$.  Both of these could affect the inferred $\sigma$ as well as the decay of averaged correlations.

We have proposed a scenario for a spin-glass phase with exponentially many states that would have definite experimental signatures. Such a phase can occur in spin-glass-like models with correlations between couplings that are special but would be hard to detect in a sample. Whether it can occur in less artificial models is unclear.  But the  scenario, if it does occur,  gives rise to many questions. What are appropriate measures for the necessary degree of frustration in couplings?  How do states change with temperature and magnetic field? What are the natures of phase transitions into such a phase?

We are grateful to Chuck Newman and Dan Stein for many fruitful discussions. This work was supported in part by the NSF via DMR0229243.


\begin{thebibliography}{4}
\bibitem{BinderYoungFisherHertz}K. Binder and A. P. Young, Rev. Mod. Phys. {\bf 58}, 801 (1986); K. H. Fischer and J. A. Hertz, {\it Spin Glasses}, vol 1 of {\it Cambridge Studies in Magnetism} (Cambridge University Press, Cambridge, 1991).

\bibitem{EA}S. F. Edwards and P. W. Anderson, J. Phys. F {\bf 5}, 965 (1975).

\bibitem{Parisi_a}G. Parisi, Phys. Rev. Lett. {\bf 43}, 1754 (1979).  See also M. M\'{e}zard, G. Parisi, and M. A. Virasoro, {\it Spin-Glass Theory and Beyond}, vol 9 of {\it Lecture Notes in Physics} (World Scientific, Singapore, 1987).

\bibitem{AndersonPondBMMcMillanFH1} P. W. Anderson and C. M. Pond, Phys. Rev. Lett. {\bf 40}, 903 (1978); A. J. Bray and M. A. Moore, J. Phys. C: Solid State Phys. {\bf 17} L463 (1984); W. L. McMillan, J. Phys. C: Solid State Phys. {\bf 17} 3179 (1984); D. S. Fisher and D. A. Huse, Phys. Rev. Lett. {\bf 56} 1601 (1986).

\bibitem{FH2}D. S. Fisher and D. A. Huse, Phys. Rev. B {\bf 38}, 386 (1988).

\bibitem{FH_many_states}Daniel S Fisher and David A Huse, J. Phys. A: Math. Gen. {\bf 20}, L1005 (1987).

\bibitem{Thouless}By contrast, the number of stable states in infinite range models or models on Bethe lattices can grow exponentially with volume. See e.g. D. J. Thouless, Phys. Rev. Lett. {\bf 56}, 1082 (1986).

\bibitem{RuelleSinai}D. Ruelle, {\it Thermodynamic Formalism} (Addison-Wesley, Reading, MA, 1978); Y. G. Sinai, {\it Theory of Phase Transitions} (Pergamon, Oxford, 1982).

\bibitem{NS4_groundstates} C.M. Newman and D.L. Stein, Phys. Rev. E {\bf 55}, 5194 (1997).

\bibitem{Middleton1}A.A. Middleton, Phys. Rev. Lett. {\bf 83}, 1672 (1999).

\bibitem{NS1_groundstates}C.M. Newman and D.L. Stein, Phys. Rev. B {\bf 46}, 973 (1992).

\bibitem{ellscale} How small $\ell$ must be can be subtle:  there may be more than one characteristic growing  length scale.  O. L. White and D. S. Fisher in preparation.

\bibitem{NS5_groundstates}C.M. Newman and D.L. Stein, Phys. Rev. E {\bf 57}, 1356 (1998).

\bibitem{ImryMa} Y. Imry and S-K. Ma, Phys. Rev. Lett. {\bf 35} 1399 (1975).

\bibitem{Imbrie} Even for arbitrarily small inter-cluster coupling, some sections of clusters will break off because of anomalously strong couplings to other clusters. But large broken-off sections should be  exponentially rare and will renormalize the original clusters only weakly.  See related discussion in J. Z. Imbrie, Phys. Rev. Lett. {\bf 53}, 1747 (1984).

\bibitem{NS1_highly_disordered}C.M. Newman and D.L. Stein, Phys. Rev. Lett. {\bf 72}, 2286 (1994).

\bibitem{Pemantle1Benjamini1}R. Pemantle, Ann. Probab. {\bf 19}, 1559 (1991); I. Benjamini, R. Lyons, Y. Peres, and O. Schramm, Ann. Probab. {\bf 29}, 1 (2001).

\bibitem{Pelcovits}R. A. Pelcovits, E. Pytte, and J. Rudnick, Phys. Rev. Lett. {\bf 40}, 476 (1978);

\bibitem{FisherXY}
D. S. Fisher, Phys. Rev. Lett. {\bf 78}, 1964 (1997).

\bibitem{incip-perc} Incipient percolation clusters are  neither space filling nor sufficiently well-connected; at any scale they can be broken in two by cutting  $\mathcal{O}(1)$ bonds. 
\bibitem{footnote} This hierarchical structure illustrates a difficulty with renormalization group (RG) descriptions: from one scale to the next the number of coarse-grained degrees of freedom is reduced, but the number that need to be retained still grows exponentially with the length scale.

\end{thebibliography}
\end{document}